# Element Abundances in Solar Energetic Particles and the Solar Corona

**Donald V. Reames**

Institute for Physical Science and Technology

University of Maryland, College Park, MD 20742-2431 USA

email: dvreames@umd.edu

**Abstract** This is a study of abundances of the elements He, C, N, O, Ne, Mg, Si, S, Ar, Ca, and Fe in solar energetic particles (SEPs) in the 2 – 15 MeV amu$^{-1}$ region measured on the *Wind* spacecraft during 54 large SEP events occurring between November 1994 and June 2012. The origin of most of the temporal and spatial variations in abundances of the heavier elements lies in rigidity-dependent scattering during transport of the particles away from the site of acceleration at shock waves driven out from the Sun by coronal mass ejections (CMEs). Variation in the abundance of Fe is correlated with the Fe spectral index, as expected from scattering theory but not previously noted. Clustering of Fe abundances during the "reservoir" period, late in SEP events, is also newly reported. Transport-induced enhancements in one region are balanced by depletions in another, thus, averaging over these variations produces SEP abundances that are energy independent, confirms previous SEP abundances in this energy region, and provides a credible measure of element abundances in the solar corona. These SEP-determined coronal abundances differ from those in the solar photosphere by a well-known function that depends upon the first ionization potential (FIP) or ionization time of the element.

## 1. Introduction

Ever since the pioneering work of Meyer (1985) it has been known that the abundances of elements in large "gradual" solar energetic particle (SEP) events are related to the corresponding element abundances in the solar corona. SEPs in gradual events are accelerated in proportion to the ambient "seed population" by shock waves, driven out from the Sun by coronal mass ejections (CMEs), beginning at from two to three solar radii (*e.g.* Reames, 1990, 1995b, 1999, 2002, 2009a, b, 2013; Kahler, 1992, 1994, 2001; Gosling, 1993; Lee 1997, 2005; Tylka, 2001; Gopalswamy *et al.*, 2002, Ng, Reames, and Tylka, 2003; Tylka, *et al.*, 2005; Tylka and Lee, 2006; Cliver and Ling 2007; Ng and Reames, 2008; Sandroos and Vainio, 2009; Rouillard *et al.*, 2011, 2012).  These gradual SEP events differ from smaller "impulsive" SEP events that are distinguished by 1000-fold enhancements of $^3$He/$^4$He and of heavy elements, *e.g.* (Z>50)/O, produced during acceleration by resonant wave-particle interactions in solar flares and jets (for a recent review of gradual and impulsive SEP events see Reames, 2013)

Meyer (1985) found, that when the small $^3$He-rich impulsive events were excluded, the element abundances in the large SEP events were composed of i) a "mass-dependent" factor that could be ordered by the abundance ratio Fe/O that he found was actually based upon the mass-to-charge ratio, *A/Q*, of the ions and ii) a pattern where the ratio of the SEP abundance to the corresponding photospheric abundance of each element depended in a simple way upon the first ionization potential (FIP) of the element.  Mainly, elements with a low FIP (<10 eV) were enhanced by a factor of about four relative to those with high FIP.  Since the accelerated ions were observed to be highly ionized, the FIP-dependent fractionation was believed to occur much earlier, during transport of low-FIP ions and high-FIP neutral atoms from the photosphere to the corona.  Thus the FIP-dependent SEP abundances were a measure of element abundances in the solar corona.

At nearly the same time, Breneman and Stone (1985), using recent charge measurements of Luhn *et al.* (1984), found that SEP abundances followed a power law *vs*. *Q/A* that was rising for some events and falling for others.  Later, Reames (1995a), using new SEP abundance measurements, found that simply averaging the abundances over a large number of events seemed to remove the *Q/A* dependence, *i.e. Q/A* enhancements in some events balanced the depletions in others.  This averaging left the same FIP enhancement for species such as Mg and Fe, which have similar FIP but much different values of *Q/A*.   The Reames (1995a) abundances have become somewhat of a reference standard of SEP abundances (see also Reames, 1998, 1999).

The clean distinction of the abundances in gradual and impulsive events was called into question when Mason, Mazur, and Dwyer (1999) found enhancements of $^3$He in some of the large gradual events that were supposed to have only coronal abundances.  The value of $^3$He/$^4$He = (1.9 ± 0.2)×10$^{-3}$ was small compared with values > 0.1 in impulsive events, but was still about five times the abundance in the solar wind.  The authors suggested that residual suprathermal ions, left over from earlier impulsive flares, were contributing to the seed population accelerated by the shock. Later, Tylka *et al.* (2005) found striking differences in the energy dependence of Fe/C, above about 10 MeV amu$^{-1}$, in otherwise similar SEP events. The power-law spectra of shock-accelerated particles suddenly steepen at high energies because of factors such as the finite acceleration time (Ellison and Ramaty, 1985; Lee, 2005). These spectral "knees" are often approximated as an exponential at an e-folding energy that depends upon $Q/A$ and upon sec $\theta_{Bn}$, where $\theta_{Bn}$ is the angle between the magnetic field vector, ***B***, and the shock normal, to account for the more efficient acceleration at quasi-perpendicular shock waves (Lee, 2005; Tylka and Lee, 2006). Since *Q* also depends upon the source of the suprathermal ions, large variations can occur above about ten MeV amu$^{-1}$, and any attempt to determine the underlying coronal abundances is best focused on the one to ten MeV amu$^{-1}$ region (see Reames, 2013). With this new understanding, are the abundances of Reames (1995a) still valid?

Extensive measurements of SEP abundances have been made at energies below about one MeV amu$^{-1}$ and above about ten MeV amu$^{-1}$ (*e.g.* Desai *et al.*, 2003, 2004, 2006; Slocum *et al.*, 2003; Cohen *et al.*, 2005, 2007; Tylka *et al.*, 2005, 2006; Leske *et al.*, 2007; Reames and Ng, 2010). However, different energy regions involve different physical processes and are not directly comparable. At energies below one MeV amu$^{-1}$, in very large SEP events, intensities are strongly suppressed until the approach of the shock near one AU (*e.g.* Desai *et al.*, 2006), so that abundances are more typical of accelerated local plasma (Desai *et al.*, 2006) than of the corona. This low-energy suppression probably results from the "streaming limit" (*e.g.* Reames and Ng, 2010) when intense streaming protons of above about 10 MeV generate Alfvén waves that scatter and retard ions coming behind; these waves strongly suppress the spectrum of ions near one MeV amu$^{-1}$ and below until arrival of the shock itself. Above ten MeV amu$^{-1}$, Tylka *et al.* (2005) found two otherwise identical SEP events where Fe/C rose sharply in one event and fell precipitously in the other, differing by a factor of about 100 near 60 MeV amu$^{-1}$. This divergence presumably resulted from differences in the energy spectra produced by different geometry at the accelerating shock (Tylka and Lee, 2006). Thus we leave the study of higher and

.

lower energies to others and limit the focus of this article to the physics of ions in the region from about 2 to 10 MeV amu$^{-1}$.

As noted above, a problem with the use of SEP observations to determine coronal abundances is that particle scattering during transport from source to observer alters the abundances and energy spectra of the ions. We will see that the scattering mean free path depends upon a small power of a particle's magnetic rigidity, or momentum per unit charge (Parker, 1963; Ng, Reames, and Tylka, 2003; Tylka *et al.*, 2012). This makes scattering depend upon both *A/Q* and velocity and is a principal cause of the abundance variations observed by Meyer (1985) and by Breneman and Stone (1985). Fe scatters less than C or O at constant velocity, for example, and high-energy particles less than those at low energy, so we see relatively Fe-rich SEPs with hard energy spectra early in events and Fe-poor SEPs with softer spectra later. Tylka *et al.* (2012) recently studied this by comparing the time dependence of Fe/O viewed from different spatial locations using *Wind* and *Ulysses*. To successfully recover the abundances at the acceleration source we must understand how transport has distributed all of the particles in both space and time, *e.g.* we must look at time dependences with observations that view many SEP events from different solar longitudes.

The *Low Energy Matrix Telescope* (LEMT: von Rosenvinge *et al.*, 1995) on the *Wind* spacecraft measures elements from He through about Pb in the energy region about 2 – 20 MeV amu$^{-1}$, identifying and binning the major elements from He to Fe onboard at a rate up to about $10^4$ particles s$^{-1}$. Instrument resolution and aspects of the processing have been shown and described elsewhere (Reames *et al.*, 1997; Reames, Ng, and Berdichevsky, 2001; Reames, 2000; Reames and Ng, 2004). Abundances of the elements, measured over a large number of gradual SEP events using LEMT, have not been reported previously.

## 2. Motivation and Event Selection

A first consideration is the elimination of impulsive SEP events, with their extreme abundances, from our event sample. Both classes of events have enhancements that depend upon *A/Q*, but the behavior is quite different, as shown in Figure 1, because of differences in underlying mechanisms and charge states.

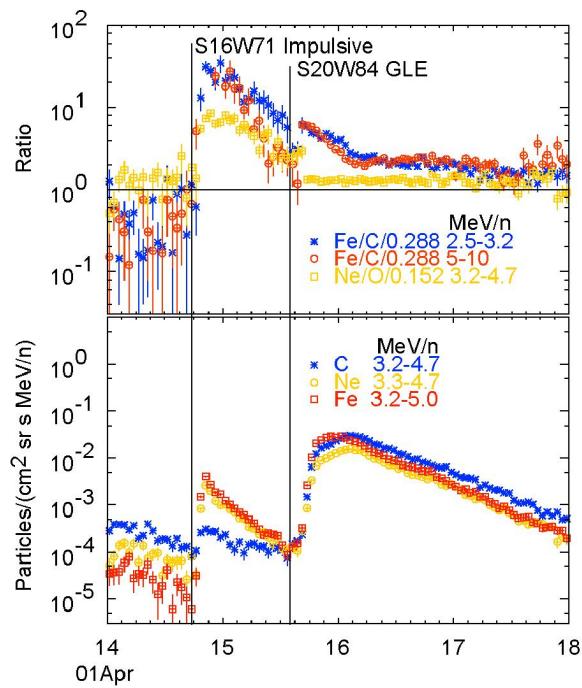

**Figure 1**. Intensities of C, Ne, and Fe *vs.* time are shown in the lower panel while ratios of Fe/C at two energies and Ne/O at one, normalized to the reference SEP abundances of Reames (1995a), are shown in the upper panel. An impulsive SEP event at solar longitude W71 begins on 14 April and a large gradual event at W84 on 15 April. The events have comparable enhancements in Fe/C, elements with widely separated values of *A/Q*, but show greatly different behavior in Ne/O, closely spaced in *A/Q*. (see Tylka *et al.*, 2002)

The impulsive SEP event on 14 April 2001 is one of the largest of the 18-year observation period of *Wind*, yet the C intensity is quite low relative to the much larger gradual event. The intensity, and the event duration of less than a day, will serve to eliminate impulsive events from our sample.

As noted above, a principle cause of the Fe/C enhancement in gradual events is different transport of Fe and C to the observer. For particle scattering against a Kolmogorov spectrum of Alfvén waves, for example, quasi-linear theory predicts the pitch-angle diffusion coefficient $D_{\mu\mu} \sim vP^{-1/3}$, where $v$ is the particle velocity and the magnetic rigidity $P = pc/Qe$ is the momentum per unit charge of the SEP ion. This implies that the scattering mean free path, $\lambda \sim v/ D_{\mu\mu} \sim P^{1/3}$ (Parker, 1963; Ng, Reames, and Tylka, 2003). Comparing ions of the same velocity, $\lambda \sim (A/Q)^{1/3}$, so that Fe with $A/Q \approx 4 - 5$, typically twice that of C, would have a 25 – 40 % greater value of $\lambda$.  This means that Fe propagates out ahead of C producing an Fe-rich region early in an SEP event and leaving an Fe-depleted region behind (see also Tylka *et al.*, 2012).  Other ions evolve similarly in *A/Q* order.

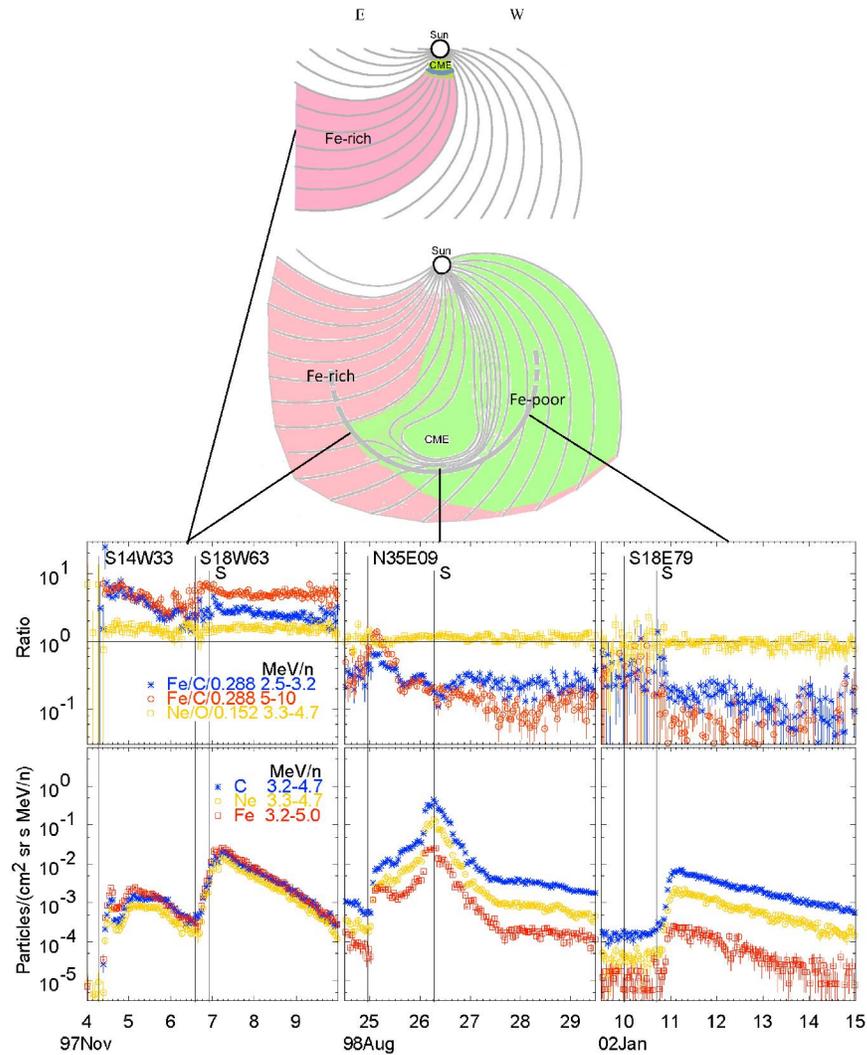

**Figure 2**.  The lower panels show intensities (below) and relative abundance ratios (above) for four different SEP events viewing sources at the solar longitudes specified.  S denotes a time of shock passage.  Events viewed from the East of the source see Fe/C enhancements while those at central and western longitudes see Fe/O depletions.  Ne/O variations are minimal.  Cartoons above show spatial evolution of a CME and shock from which Fe escapes more easily than C (see text).

The SEP events in the left panels of Figure 2, and the large event in Figure 1, show the typical enhancement of Fe/C for events on the East flank of the shock (*i.e.* a western source on the Sun).  These events are magnetically well-connected to the shock when it is near the Sun and the Parker spiral of the field sweeps the Fe-rich ions, dominating initially, toward the East.  Subsequently, the shock itself propagates through and reaccelerates the Fe-depleted material left over from the earlier acceleration. Incidentally, impulsive events have *A/Q*-dependent enhancements in the source flares *in addition* to these effects of transport.

*If we wish to recover the source coronal abundances we must, at a minimum, correctly average over the temporal and spatial distribution induced by differential transport*. This corresponds to the "mass-unbiased" abundances of Meyer (1985) or the averaged abundances of Reames (1995a). Other *A/Q*-dependent enhancements, such as the impulsive-suprathermal ions in the seed populations (Desai *et al.*, 2003, 2006; Tylka *et al.*, 2005; Tylka and Lee, 2006), must also be considered, although their effects are greatest above ten MeV amu$^{-1}$.

The elemental resolution of the LEMT telescope is shown in Figure 3. Although resolution is decreased at the lowest energies, it is still adequate for study of the energy dependence of abundances of the major species.

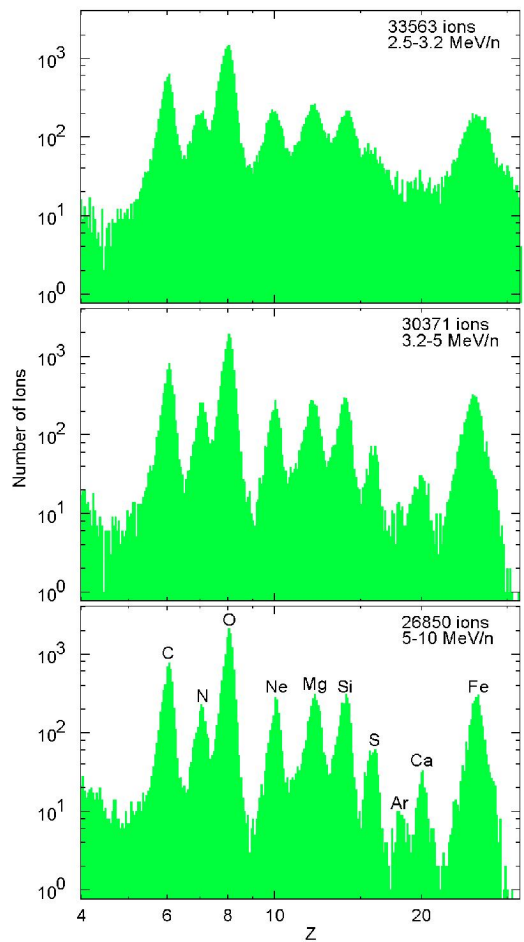

**Figure 3**. Element resolution of the LEMT telescope for 4 < Z < 30 is shown in three energy intervals. The number of ions included in each histogram is also shown. Elements resolved in the highest energy interval are indicated in the lower panel. The resolution improves with increasing energy.

Elemental resolution of the instrument used by Reames (1995a) on the *International Sun-Earth Explorer-3* (ISEE 3) was superior to that of LEMT, only partly because the former provided pulse-heights from three detectors for each particle above five MeV amu$^{-1}$ while LEMT has only two. This redundancy also allowed better background rejection at high intensities. However, LEMT, with about 100 times the geometrical factor and with onboard binning, can measure as many ions in a few hours of a large SEP event as ISEE 3 did in its seven-year mission. This makes it easy for us to study the temporal evolution of abundances within each large gradual SEP event along with the event-to-event variations.

Most previous abundance studies have been limited to event averages. We begin our study with eight-hour intervals, choosing those that have sufficient intensities to exceed the intensities of impulsive events and to provide a significant sample for abundance studies of all major elements at all energies. The eight-hour periods provide enough time resolution to follow the trend of abundance variations within an event while maintaining a manageable number of points to also show a comparison of many events. In Figure 4 we show a cross-plot of the intensities of Fe and O at 3.2 – 5 MeV amu$^{-1}$ for the 387 candidate eight-hour intervals from 4 November 1994 to 14 June 2012.

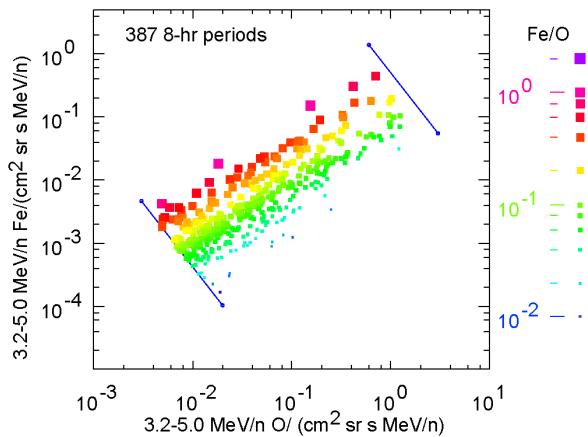

**Figure 4**. Intensity of Fe *vs*. O at 3.2 – 5 MeV amu$^{-1}$ is shown for candidate eight-hour intervals. The diagonal boundaries shown are orthogonal to the trend of the data. The upper bound is required to eliminate 12 intense periods of high background and reduced resolution.

Unfortunately, it was necessary to remove 12 eight-hour periods at high intensities because of instrument saturation effects that increased background and reduced the resolution. Finally we grouped the eight-hour periods into SEP events, each with a well defined source CME, also removing events with less than three eight-hour periods. This resulted in 54 SEP events involving 342 eight-hour periods.

## 3. Time Dependence of the Abundances

We begin our study by examining abundance variations in the single intermediate energy interval 3.2 – 5 MeV amu$^{-1}$. Figure 5 shows the distribution of the points in the space of Si/O *vs.* Fe/C, normalized to the corresponding SEP abundances of Reames (1995a). Different denominators, C and O, are chosen to avoid possible coordinated variation caused by a repeated denominator. The variations in Si/O and Fe/C are strongly correlated although Si/O varies by a factor of only about two while Fe/C varies by a factor of about ten. Note that the width of the distribution, orthogonal to the principal axis of variation, is actually quite narrow. The correlated behavior of Si/O *vs.* Fe/C is expected from the *A/Q* dependence of the transport discussed above.

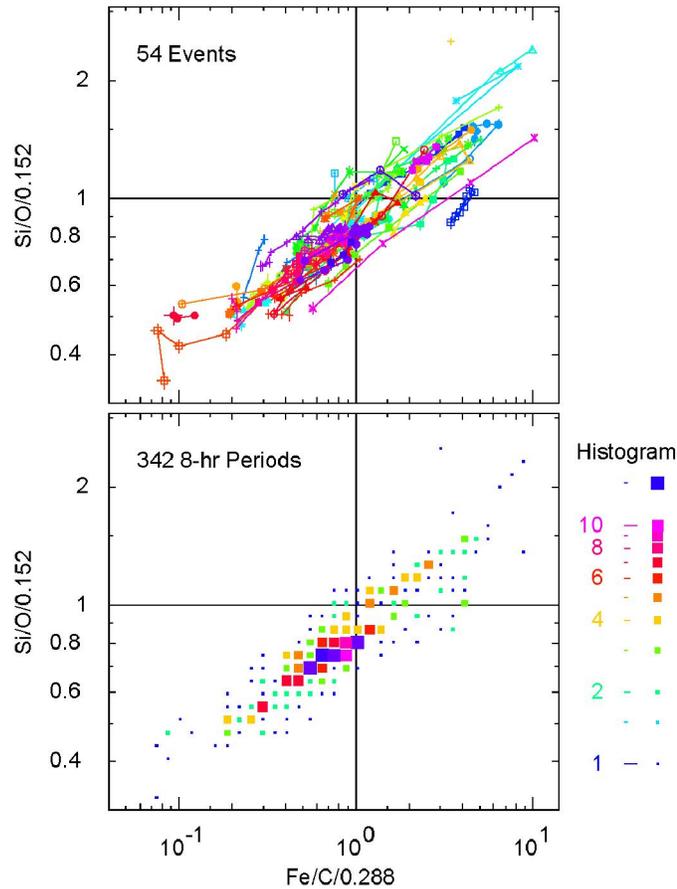

**Figure 5**. The upper panel shows the pattern of the eight-hour periods grouped by color and symbol into the 54 SEP events in a space of Si/O *vs.* Fe/C, at 3.2 – 5 MeV amu$^{-1}$, normalized to abundances from Reames (1995a). Points tend to proceed from the upper right toward the lower left with time during most events. The lower panel is a histogram of the same data, showing a strong clustering of points, many occurring in the depleted "reservoir" region late in SEP events.

The histogram in Figure 5 shows a strong clustering of points in an Fe- and Si-depleted region. This is the "reservoir" region (Reames, 2013) behind the shocks for many events, although some events begin and end here. The temporal dependence of this behavior is illustrated in Figure 6 showing the daily evolution of the distribution of the sequence numbers of eight-hour periods within an event in the Si/O *vs.* Fe/C space. Eleven of the events persist to contribute to days 4 or 5. It is surprising that all 11 events tend to converge to the same depleted value of Fe/C.

**Figure 6**. Panels show the evolution of the clustering in the normalized Si/O *vs.* Fe/C space at 3.2 – 5 MeV amu$^{-1}$ on successive days within an event. The small numbers are the sequence numbers of the eight-hour periods within an SEP event (1, 2, and 3 are in the first day of an event, 4, 5, and 6 in the second, *etc.*).

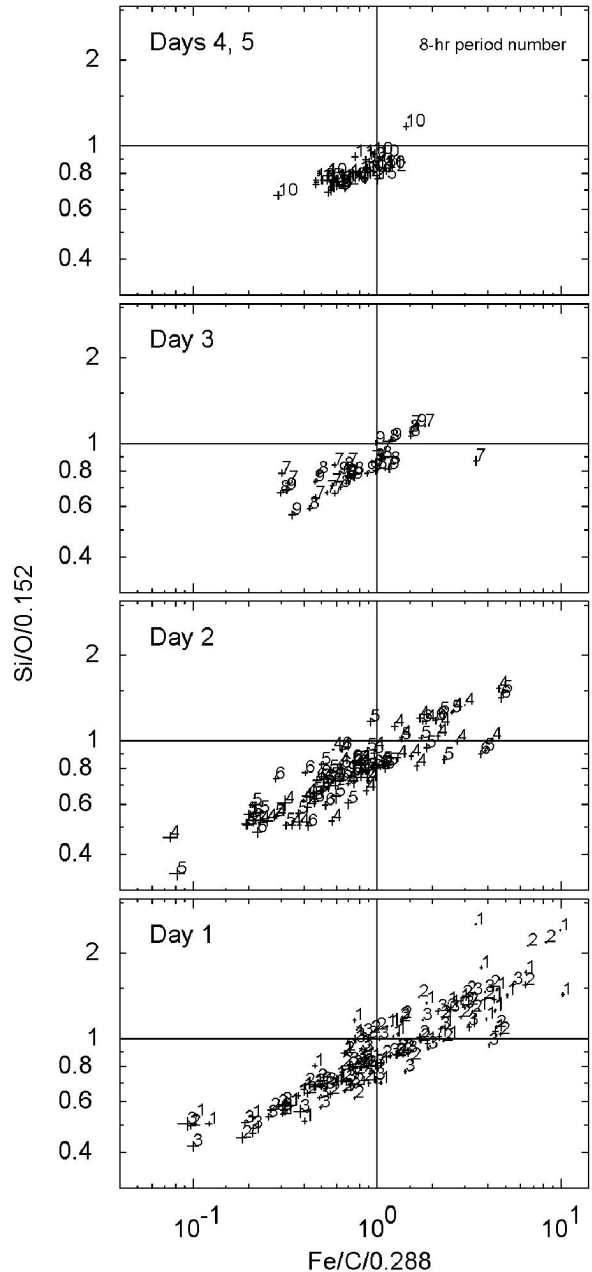

Element Abundances in SEPs

For abundance ratios that involve elements with more-similar values of *A/Q*, we would not expect a correlation with Fe/C. Examples of this are shown in Figure 7 for normalized values of He/O and Ne/O *vs*. Fe/C.

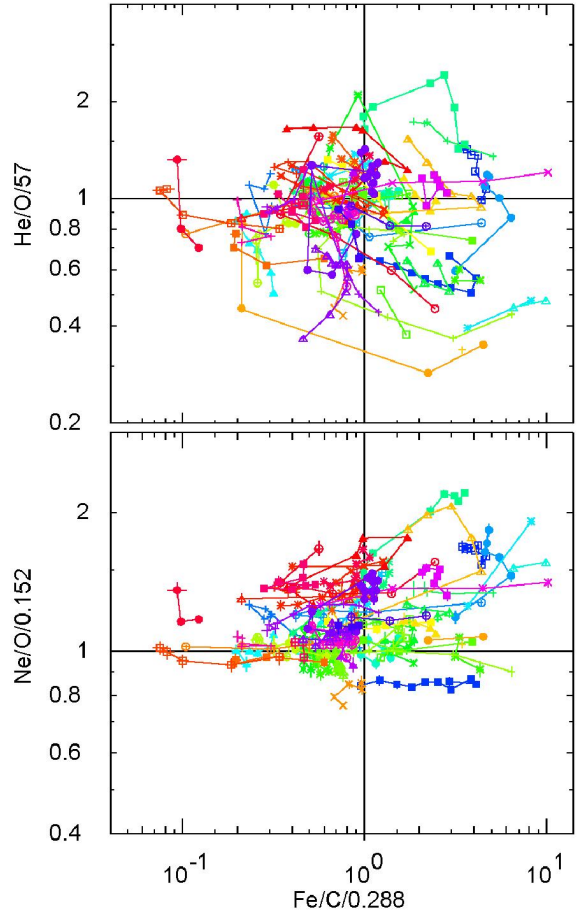

**Figure 7**. Upper and lower panels show the normalized values of He/O and Ne/O, respectively, *vs*. Fe/C at 3.2 – 5 MeV amu$^{-1}$ for the 342 eight-hour intervals grouped by color and symbol into 54 SEP events. The variations are uncorrelated, unlike those in Si/O shown in the upper panel of Figure 5, primarily because He, O, and Ne have similar values of *A/Q* while Fe and C do not. In most cases, points move from right to left during an event.

## 4. Energy Spectral Correlations

We study energy spectral effects after averaging each available energy interval over all eight-hour time intervals within each SEP event, effectively summing the ions measured in each energy interval so that Fe-rich and Fe-poor times during each event are allowed to partially compensate. For each of the major species, He, C, O, Ne, Si, and Fe, six or seven energy intervals divide the 2 – 15 MeV amu$^{-1}$ study region. Power-law fits to these energy spectra can be obtained for each species in each of the 54 SEP events to compare spectra of different species and possible correlations between spectra and abundances. The spectral fits are generally quite good, and their relative quality in indicated by their error bars.

D.V. Reames

Figure 8 shows a cross-plot of the spectral indices of He and Fe *vs.* that of O obtained in a common 2.5 – 10 MeV amu$^{-1}$ region for each of the 54 events. *A priori*, one might have greater confidence in the abundances if the spectral indices of all species were the same in each SEP event. However, the same transport that produces an *A/Q* dependence for different species at constant velocity can also affect the rigidity or energy dependence of the spectra of each species. The large scatter in the spectra of Fe and O is of the same origin as the variations in Fe/O. The ion velocity and the rigidity dependence of the scattering allow high-energy ions to move outward preferentially, producing flatter spectra early and steeper spectra in the depletion region behind. We might expect this to produce a spatial pattern of spectral indices similar to that for Fe/O shown in the upper panels of Figure 2. However, the origin of the weak systematic tendency for He and Fe spectra to be slightly harder than the O spectra for steeper spectra is not clear.

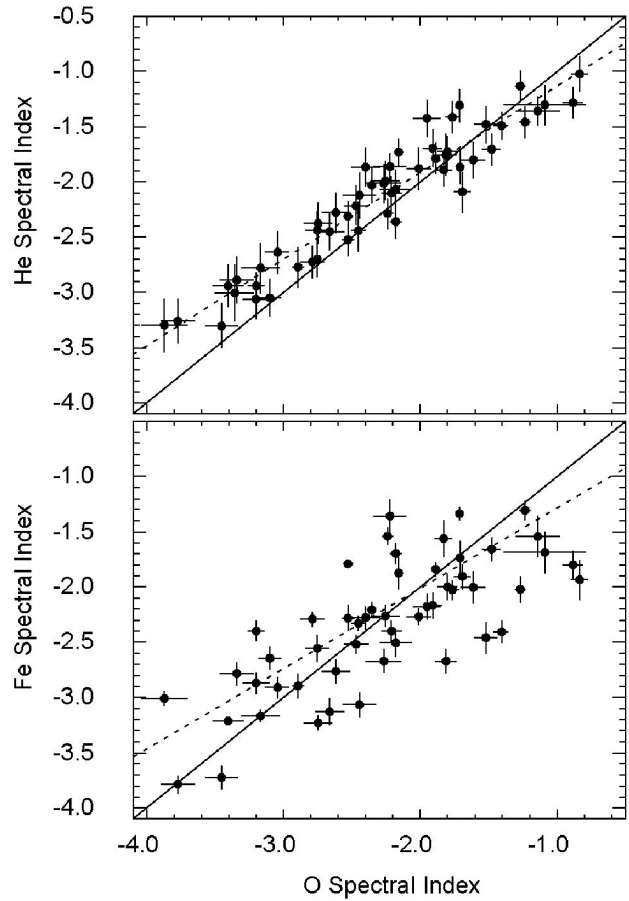

**Figure 8**. Upper and lower panels show the energy spectral indices of He and Fe, respectively, *vs.* that of O, over the 2.5 – 10 MeV amu$^{-1}$ interval, for each of the 54 SEP events studied. Error bars show errors in the least-squares power-law fits to the energy spectra. Diagonal lines are solid in each panel and weighted least-squares fits to each distribution of spectral indices are shown dashed.

We explore possible correlations between abundances and spectral indices in Figure 9. Here there is a strong correlation between abundance and spectral index for Fe shown in the lower panel of the figure that does not exist for He or Ne in the panels above.

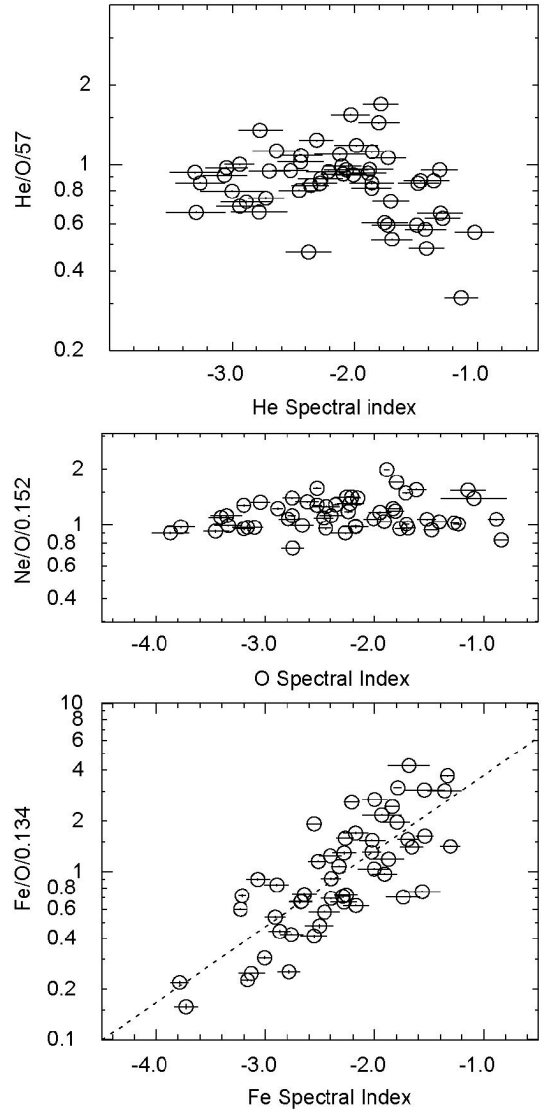

**Figure 9**. The relationship between abundances and spectral indices for SEP events are shown for He (upper panel), Ne (middle panel) and Fe (lower panel). A least-squares fit line is shown in the lower panel which has a correlation coefficient of 0.79.

## 5. The Average SEP Abundances

We use un-weighted averaging over the 54 SEP events to derive averaged abundances at each energy. The number of particles in each event is sufficiently large that statistical errors are negligible and summing the *particles* over all the events would allow a few large events to dominate the averages completely. While the averaging method is unimportant for Ne/O, for example, it does matter for Fe/O. The smaller SEP events help us average over the spatial variations.

Results for the energy variations for the dominant elements are shown in Figure 10 and compared with the abundances of Reames (1995a).

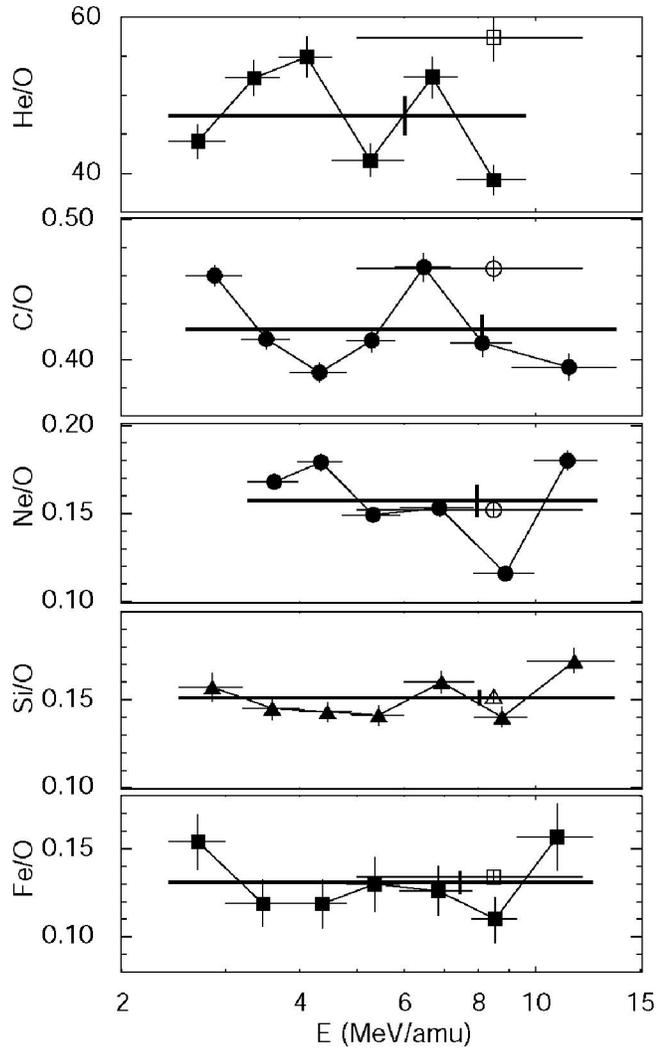

**Figure 10**. Element abundances, averaged over the SEP events, are shown as a function of energy as solid symbols and overall energy averages as solid bars for this work. Open symbols show the corresponding abundances from Reames (1995a).

The data in Figure 10 show little evidence for any systematic energy dependence. This suggests that our averaging has compensated for much of the energy- and $A/Q$-dependent spreading in space and time.

To illustrate the importance of balancing the Fe-rich and Fe-poor regions of an SEP event we return to reconsider Fe/O in the reservoir region discussed earlier. The upper panel of Figure 6 shows the clustering of Si/O *vs.* Fe/O during 35 eight-hour periods on days 4 and 5 of 11 large SEP events. The reduced variation provides a well-defined Fe/O value that is surprisingly similar for all events in this 3.2 – 5 MeV amu$^{-1}$. Other energy regions show a similar clustering

Element Abundances in SEPs

of the data. However, in Figure 11 we see the energy dependence resulting from an un-weighted average over the available data in this reservoir region.

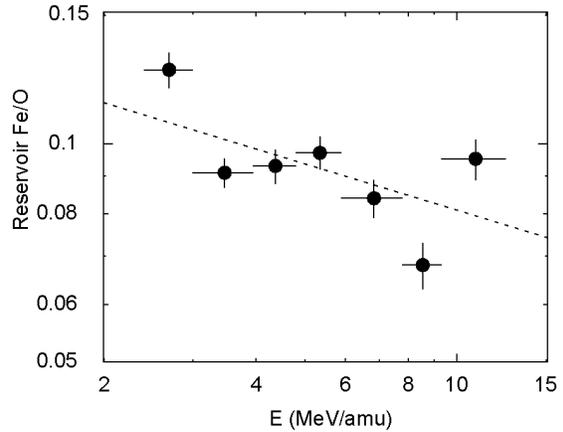

**Figure 11**. The energy dependence of Fe/O, averaged over the available data in the reservoir period on days 4 and 5 of 11 SEP events, shows a preferential depletion of Fe/O especially at high energies, as might be expected from transport theory. The dashed line is a least-squares fit. Compare with the lower panel of Figure 10.

Finally, Figure 12 shows the SEP abundances averaged over energy, divided by recent photospheric abundances. Table 1 compares the complete list of SEP element abundances measured in this work with those from Reames (1995a) and with photospheric and with spectroscopic coronal abundances.

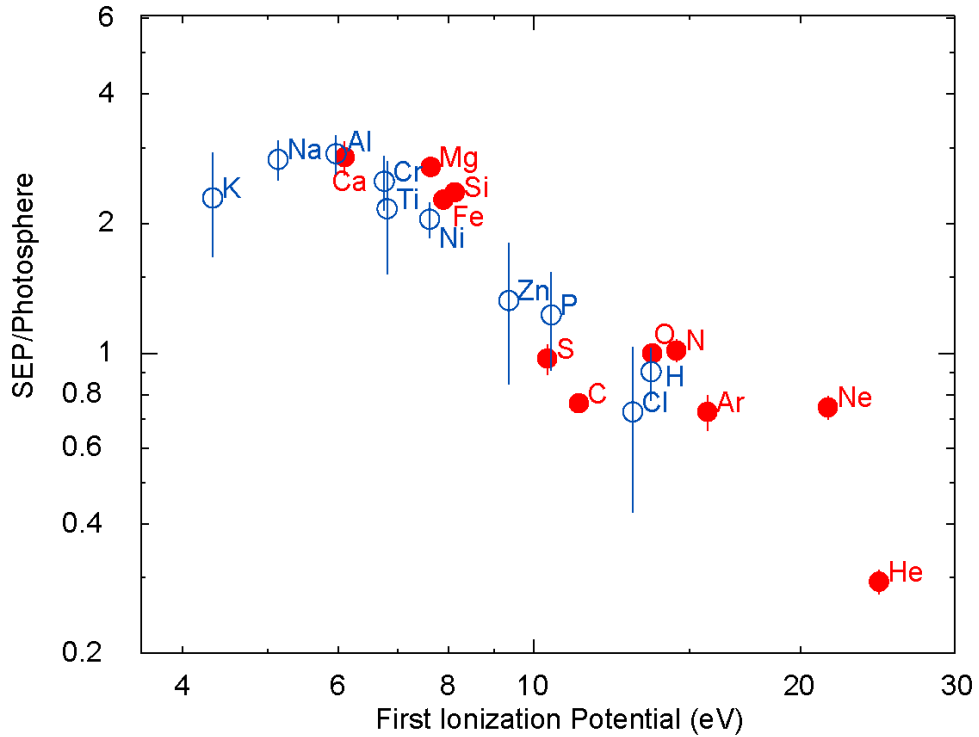

**Figure 12**. Average element abundances in SEP events relative to photospheric abundances, normalized at O, are shown as a function of FIP. Abundances from this work, shown as filled red circles, are supplemented by abundances of additional elements from Reames (1995a) shown as open blue circles. Note that errors in the photospheric abundances (about 10 – 20 %) have not been included.

D. V. Reames

Table 1 Photospheric, SEP, and Spectroscopic Coronal Abundances.

| | Z | FIP [eV] | Photosphere[1] | SEP This work | SEP[2] | Spect. Corona[3] |
|---|---|---|---|---|---|---|
| H | 1 | 13.6 | 1.74×10⁶ [*] | – | (1.57±0.22)×10⁶ | 1.50×10⁶ |
| He | 2 | 24.6 | 1.6×10⁵ | 47000±3000 | 57000±3000 | 1.28×10⁵ |
| C | 6 | 11.3 | 550±76[*] | 420±10 | 465±9 | 493 |
| N | 7 | 14.5 | 126±35[*] | 128±8 | 124±3 | 124 |
| O | 8 | 13.6 | 1000±161[*] | 1000±10 | 1000±10 | 1000 |
| Ne | 10 | 21.6 | 210 | 157±10 | 152±4 | 192 |
| Na | 11 | 5.1 | 3.68 | – | 10.4±1.1 | 11.8 |
| Mg | 12 | 7.6 | 65.6 | 178±4 | 196±4 | 225 |
| Al | 13 | 6.0 | 5.39 | – | 15.7±1.6 | 18.3 |
| Si | 14 | 8.2 | 63.7 | 151±4 | 152±4 | 215 |
| P | 15 | 10.5 | 0.529±0.046[*] | – | 0.65±0.17 | – |
| S | 16 | 10.4 | 25.1±2.9[*] | 25±2 | 31.8±0.7 | 31.8 |
| Cl | 17 | 13.0 | 0.329 | – | 0.24±0.1 | – |
| Ar | 18 | 15.8 | 5.9 | 4.3±0.4 | 3.3±0.2 | 5.77 |
| K | 19 | 4.3 | 0.224±0.046[*] | – | 0.55±0.15 | – |
| Ca | 20 | 6.1 | 3.85 | 11±1 | 10.6±0.4 | 13.2 |
| Ti | 22 | 6.8 | 0.157 | – | 0.34±0.1 | – |
| Cr | 24 | 6.8 | 0.834 | – | 2.1±0.3 | – |
| Fe | 26 | 7.9 | 57.6±8.0[*] | 131±6 | 134±4 | 187 |
| Ni | 28 | 7.6 | 3.12 | – | 6.4±0.6 | 10.5 |
| Zn | 30 | 9.4 | 0.083 | – | 0.11±0.04 | – |

[1] Lodders, Palme, and Gail (2009)

[*] Caffau *et al.* (2011)

[2] Reames (1995a)

[3] Feldman and Widing (2003)

      The published photospheric abundances of elements have been changing in recent years (see Schmelz *et al.*, 2012), largely because of improvements in modeling of the solar atmosphere. Thus as indicated in Table 1, the photospheric abundance measurements of C, N, O, P, S, K, and Fe, relative to H, were taken from Caffau *et al.* (2011). These were supplemented, for other elements, by the "solar system" abundances based mainly upon meteorites (Lodders, Palme, and

Gail, 2009). The abundances listed as spectroscopic corona in the table are the atomic spectroscopic abundances at temperatures above $1.4\times10^6$ K listed in Feldman and Widing (2003); at temperatures below $8\times10^5$ K the abundances of the low-FIP elements Na, Mg, Al, Si, Ca, Fe, and Ni are reduced by a factor of exactly two with no change in the other species, according to Feldman and Widing (2003). The listed spectroscopic abundances show a stronger FIP effect than the SEPs, while solar wind abundances (not shown, see *e.g.* von Steiger *et al.*, 2000) show a weaker FIP effect than SEPs. Each set of abundances was normalized to O for comparison with the SEP abundances since we do not measure H in this work and O is a reliable basis for SEPs. Coronal abundances from SEPs, the solar wind, and atomic spectral lines, all relative to H, were recently combined by Schmelz *et al.* (2012) to determine the average levels of the coronal high- and low-FIP abundances.

## 6. Discussion

We have described particle transport as the main source of *A/Q*-dependent variations of abundances in time and space, causing the "mass-dependent" variations of Meyer (1985) and the *A/Q*-dependent variations of Breneman and Stone (1985). Transport causes enhancements in ratios such as Fe/C (or Fe/O) for ions escaping early in SEP events and a depletion of the ions left behind, thus the abundances can either increase or decrease with *A/Q* depending on where and when we measure them (see Figure 2). If we could just perform the proper averaging, the transport effects would average out. We have shown that the rigidity or energy spectrum of a species such as Fe behaves similarly to its abundances during transport. Thus a measure of the proper average over the space-time variations is the energy dependence of the abundance ratio. The energy dependence will tend to vanish when we have the right balance of enhanced and depleted samples.

The earlier ISEE 3 measurements of Reames (1995a) might appear to use a different weighting of events since all particles sent by telemetry to the ground were included. *A priori*, this would appear to overweight high intensities and emphasize a few large events with western sources. However, the telemetry rate was severely limited for ISEE 3 so that effectively most of the SEP events were sampled equally, as they were, by design, in the present work.

D. V. ReamesThe observations of Mason, Mazur, and Dwyer (1999), Desai *et al.* (2003), and Tylka *et al.* (2005) and the theory of Tylka and Lee (2006) have emphasized the importance of impulsive suprathermal ions in the seed population for shock acceleration. These impulsive suprathermal ions, with abundance enhancements also increasing with *A/Q*, especially for Fe/O (see *e.g.* Reames, 2013), could certainly disrupt the connection between SEP abundances and those in the corona. However, our observed correlation of Fe enhancement and depletion with the Fe spectral index (Figure 9) strongly suggests that most of the variation we observe occurs *after* acceleration and thus results from transport which affects both abundances and spectra in a well-described complementary way. The strongest effects of the seed population occur above the spectral knee energies which depend upon both *Q/A* and sec $\theta_{Bn}$ (Tylka and Lee, 2006). A percentage of impulsive suprathermals, which may be small in our energy region, can have a large effect above ten MeV amu$^{-1}$ where spectral knees become dominant.

As pointed out in Figure 1, for example, impulsive-flare accelerated ions can be distinguished by Ne/O, which is minimally affected by transport. We cannot rule out the possibility of suprathermal Ne and O contributing to the event-to-event variations of Ne/O shown in the middle panel of Figure 9. The variation of an individual event from the mean for Ne/O at 3.2 – 5 MeV amu$^{-1}$ is 19.5 %. In fact, the large gradual SEP event of 2001 April 15 event shown in Figure 1, following the impulsive event, has a value of Ne/O = 0.196 ± 0.002 (statistical error) and also has a high value of Fe/O = 0.349 ± 0.002. However, this event does not have the largest value of either ratio, and it is not yet possible to correlate SEP abundances with impulsive precursors. The comparison of SEP Ne/O with the photospheric and spectroscopic coronal values in Table 1 suggests that the SEP ratio is not excessively contaminated with impulsive material or otherwise overestimated.

Another figure of merit of the SEP-to-coronal comparison is the comparison of the abundances of Mg and Si with that of Fe in Figure 12. These elements have a similar FIP but differ considerably in *A/Q*. Compare, for example the different magnitude of the *A/Q*-dependent variation of Si and Fe in Figures 5 and 6. Thus the photospheric Fe/Mg ratio is 0.88 ± 0.12 and, for this work, the SEP value is 0.74 ± 0.04.

The suppression of He stands out in Figure 12. It has been suggested that the organization of coronal abundances should be a declining power-law function of the ionization time rather than FIP (*e.g.* Marsch, von Steiger and Bochsler, 1995; Gloeckler and Geiss, 2007). A longer ionization time allows greater suppression of the abundance. On this basis He stands alone, well above Ne and O. However, the theory of coronal abundances is beyond the scope of this study.

## 7. Conclusions

Dependence of the interplanetary scattering mean free path on magnetic rigidity causes particle separation in time and space that depends upon energy, for a single species, and upon *A/Q*, for different species at the same velocity. This can cause an order-of-magnitude variation in an abundance ratio of species with factor-of-two differences in *A/Q*, such as Fe/C or Fe/O. It can also cause correlated differences in the energy-spectral index of Fe, which are observed.

Averaging over the spatial distribution of SEP events averages over regions of enhancement and depletion of Fe/O, for example, and recovers abundances of elements from He through Fe with no systematic energy dependence in the 2 – 15 MeV amu$^{-1}$ region. These abundances provide the best estimate of the coronal abundances of the elements available from these SEP measurements. Comparison with the current photospheric abundances provides a measure of the coronal fractionation or "FIP effect."

**Acknowledgments**: I would like to thank Chee Ng, and Allan Tylka for helpful discussions and comments on this manuscript.

## References

Breneman, H.H., Stone, E.C.: 1985, *Astrophys. J. Lett.* **299**, L57.

Caffau, E., Ludwig, H.-G., Steffen, M., Freytag, B., Bonofacio, P.: 2011 *Solar Phys.*, **268**, 255. doi: 10.1007/s11207-010-9541-4

Cliver, E.W., Ling, A.G.: 2007, *Astrophys. J.* **658**, 1349.

Cohen, C.M.S., Stone, E.C., Mewaldt, R.A., Leske, R.A., Cummings, A.C., Mason, G.M., Desai, M.I., von Rosenvinge, T.T., Wiedenbeck, M.E.: 2005, *J. Geophys Res. A,* **110**, A09S16.

Cohen, C.M.S., Mewaldt, R.A., Leske, R.A., Cummings, A.C., Stone, E.C., Wiedenbeck, M.E., von Rosenvinge, T.T., Mason, G.M.: 2007, *Space Sci. Rev*. **130**, 183.

D. V. ReamesDesai, M.I., Mason, G.M., Dwyer, J.R., Mazur, J.E., Gold, R.E., Krimigis, S.M., Smith, C.W., Skoug, R.M.: 2003, *Astrophys. J.* **588**, 1149.

Desai, M.I., Mason, G.M., Wiedenbeck, M.E., Cohen, C.M.S., Mazur, J.E., Dwyer, J.R., Gold, R.E., Krimigis, S.M., Hu, Q., Smith, C.W., Skoug, R. M.: 2004, *Astrophys*. *J.* **661**, 1156.

Desai, M.I., Mason, G.M., Gold, R.E., Krimigis, S.M., Cohen, C.M.S., Mewaldt, R.A., Mazur, J.E., Dwyer, J.R.: 2006 *Astrophys. J.* **649**, 740.

Ellison, D., Ramaty, R.: 1985, *Astrophys. J.* **298**, 400.

Feldman, U., Widing, K.G.: 2003, *Space Sci. Rev.* **107**, 665.

Gloeckler, G., Geiss, J.: 2007, *Space Sci. Rev.* **130**, 139.

Gosling, J.T.: 1993, *J. Geophys. Res.* **98**, 18937.

Gopalswamy, N., Yashiro, S., Michalek, G., Kaiser, M.L., Howard, R.A., Reames, D.V., Leske, R., von Rosenvinge, T.: 2002, *Astrophys. J. Lett.* **572**, L103.

Kahler, S.W.: 1992, *Ann. Rev. Astron. Astrophys*. **30**, 113.

Kahler, S.W.: 1994, *Astrophys. J.* **428**, 837.

Kahler, S.W.: 2001, *J. Geophys. Res.* **106**, 20947.

Lee, M.A.: 1997, in Crooker, N., Jocelyn, J.A., Feynman, J. (eds.) *Coronal Mass Ejections*, Geophys. Monograph 99, AGU, 227.

Lee, M.A.: 2005, *Astrophys. J. Suppl.*, **158**, 38.

Leske, R.A., Mewaldt, R.A., Cohen, C.M.S., Cummings, A.C., Stone, E.C., Wiedenbeck, M.E., von Rosenvinge, T.T.: 2007 *Space Sci. Rev.* **130**, 195.

Lodders, K., Palme, H., Gail, H.-P. 2009, in Trümper, J.E. (ed.) *Landolt-Börnstein*, New Series, vol. VI/4B (Springer, Berlin), chap. 4.4, 560.

Luhn, A., Klecker B., Hovestadt, D., Gloeckler, G., Ipavich, F. M., Scholer, M., Fan, C.Y., Fisk, L.A.: 1984, *Adv. Space Res*. **4**, 161.

Marsch, E., von Steiger, R., Bochsler, P.: 1995, Astron. Astrophys. **301**, 261.

Meyer, J. P.: 1985, *Astrophys. J. Suppl* **57**, 151.

Mason, G.M., Mazur, J.E., Dwyer, J.R.: 1999, *Astrophys. J. Lett.* **525**, L133.

Ng, C.K., Reames, D.V.: 2008 *Astrophys. J. Lett.* **686**, L123.

Ng, C.K., Reames, D.V., Tylka, A.J.: 2003, *Astrophys. J.* **591**, 461.

Parker, E.N.: 1963, *Interplanetary Dynamical Processes*: Interscience.

Reames, D.V.: 1990, *Astrophys. J. Suppl.* **73**, 235.

Reames, D.V.: 1995a, *Adv. Space Res*. **15** (7), 41.

Reames, D.V.: 1995b, *Revs. Geophys. (Suppl.)* **33**, 585.

Reames, D.V.: 1998, *Space Sci. Rev.* **85**, 327.

Reames, D.V.: 1999, *Space Sci. Rev.,* **90**, 413.

Reames, D.V.: 2000, *Astrophys. J. Lett.* **540**, L111.

Reames, D.V.: 2002, *Astrophys. J. Lett.* **571**, L63.

Reames, D.V.: 2009a, *Astrophys. J.* **693**, 812

Reames, D.V.: 2009b, *Astrophys. J.* **706**, 844

Reames, D.V.: 2013, *Space Sci. Rev.* DOI 10.1007/s11214-013-9958-9
20

Reames, D.V., Barbier, L.M., von Rosenvinge, T.T., Mason, G.M., Mazur, J.E., Dwyer, J.R.: 1997, *Astrophys. J.* **483**, 515.

Reames, D.V., Ng, C.K.: 2004, *Astrophys. J.* **610**, 510.

Reames, D.V., Ng C.K.: 2010 *Astrophys. J.* **723**, 1286.

Reames, D.V., Ng, C.K., Berdichevsky, D.: 2001, *Astrophys. J.* **550**, 1064.

Rouillard, A.C., Odstrčil, D., Sheeley, N.R. Jr., Tylka, A.J., Vourlidas, A., Mason, G., Wu, C.-C., Savani, N.P., Wood, B.E., Ng, C.K., *et al.*: 2011, *Astrophys. J.* **735**, 7.

Rouillard, A., Sheeley, N.R.Jr., Tylka, A., Vourlidas, A., Ng, C.K., Rakowski, C., Cohen, C.M.S., Mewaldt, R.A., Mason, G.M., Reames, D.,*et al.*: 2012, *Astrophys. J.* **752**:44.

Sandroos, A., Vainio, R.: 2009 A*stron. and Astrophys*. **507**, L21.

Schmelz , J.T., Reames, D.V., von Steiger, R., Basu, S.: 2012, *Astrophys. J.* **755**:33.

Slocum, P.L., *et al.* 2003, *Astrophys. J.* **594**, 592.

Tylka, A.J.: 2001, *J. Geophys. Res.* **106**, 25333.

Tylka, A.J., Boberg, P.R., Cohen, C.M.S., Dietrich, W.F., Maclennan, C.G., Mason, G.M., Ng, C.K., Reames, D.V.: 2002, *Astrophys. J. Lett.* **581**, L119.

Tylka, A.J., Cohen, C.M.S., Dietrich, W.F., Lee, M.A., Maclennan, C.G., Mewaldt, R.A., Ng, C.K., Reames, D.V.: 2005, *Astrophys. J.* **625**, 474.

Tylka, A.J., Cohen, C.M.S., Dietrich, W.F., Lee, M.A., Maclennan, C.G., Mewaldt, R.A., Ng, C.K., Reames, D.V.: 2006, *Astrophys. J. Suppl.* **164**, 536.

Tylka, A.J., Lee, M.A.: 2006, *Astrophys. J.* **646**, 1319.

Tylka, A.J., Malandraki, O.E., Dorrian, G., Ko, Y.-K., Marsden, R.G., Ng, C.K., Tranquille, C.: 2012 *Solar Phys* **172**, DOI:10.1007/s11207-012-0064-z

von Rosenvinge, T.T., Barbier, L.M., Karsch, J., Liberman, R., Madden, M.P., Nolan, T., Reames, D.V., Ryan, L., Singh, S. *et al.*: 1995, *Space Sci. Rev.* **71**, 155.

von Steiger, R., Schwadron, N.A., Fisk, L.A., Geiss, J., Gloeckler, G. Hefti, S., Wilken, B., Wimmer-Schweingruber, R.F., Zurbuchen, T.H.: 2000, *J. Geophys. Res*. **105**, 27217.